\documentclass[prb,preprint]{revtex4}

\usepackage{amsmath}
\usepackage{graphicx}

\newcommand{\grad}{\boldsymbol{\nabla}}
\newcommand{\xib}{\boldsymbol{\xi}}

\newcommand{\xv}{\mathbf{x}}
\begin{document}

\title{Multipole radiation fields from the Jefimenko
equation for the magnetic field and the
Panofsky-Phillips equation for the electric field}

\author{R. de Melo e Souza}
\email{reinaldo@if.ufrj.br}
\author{M.V. Cougo-Pinto}
\author{C. Farina}

\affiliation{Universidade Federal do Rio de Janeiro,
Instituto de Fisica, Rio de Janeiro, RJ 21945-970}

\author{M. Moriconi}
\affiliation{Universidade Federal Fluminense, Instituto de Fisica,
Niteroi, RJ 24210-340}


\begin{abstract}
We show how to obtain the first multipole contributions to the
electromagnetic radiation emited by an arbitrary localized source
directly from the Jefimenko equation for the magnetic field and the
Panofsky-Phillips equation for the electric field. This procedure
avoids the unnecessary calculation of the electromagnetic
potentials.
\end{abstract}

\maketitle

\section{Introduction}

The derivation found in most textbooks of the electromagnetic fields
generated by arbitrary sources in vacuum starts by calculating the
corresponding electromagnetic potentials (see for instance,
Refs.~\onlinecite{Griffiths}, \onlinecite {Marion}, or
\onlinecite{Jackson2nd}). After the retarded potentials are obtained
(assuming the Lorenz gauge) the electromagnetic fields are
calculated with the aid of the relations ${\bf E} = - \grad\Phi -
(1/c)(\partial{\bf A}/\partial t)$ and ${\bf B}= \grad\times{\bf A}$
(we use Gaussian units). The resulting expressions for the fields
are usually called Jefimenko's equations because they appeared for
the first time in the textbook by Jefimenko.\cite{JefimenkoBook}
Jefimenko's equations are obtained in Ref.~\onlinecite{Griffiths}
from the retarded potentials and are obtained directly from
Maxwell's equations in Ref.~\onlinecite{Jackson3nd}.
(Heras\cite{Heras95} had already derived Jefimenko's equations
directly from Maxwell's equations.) References~\onlinecite{McDonald}
and \onlinecite{PanofskyPhillips} obtain a less common form of
Jefimenko's equations for the electric field, but this form is more
convenient for studying radiation.

Griffiths and Heald\cite{GriffithsHeald91} illustrate Jefimenko's
equations by obtaining the standard Li\'enard-Wiechert fields for a
point charge. Ton\cite{TranCongTon91} provides an alternative
derivation of Jefimenko's equations and three applications,
including that of a point charge in arbitrary motion. Heras has
generalized Jefimenko's equations to include magnetic monopoles and
obtained the electric and magnetic fields of a particle with both
electric and magnetic charge in arbitrary motion.\cite{Heras94A} He
has also discussed Jefimenko's equations in material media to obtain
the electric and magnetic fields of a dipole in arbritary
motion\cite{Heras94B} and has derived Jefimenko's equations from
Maxwell's equations using the retarded Green function of the wave
equation.\cite{Heras95}

The main purpose of this paper is to enlarge the list of problems
that are solved directly from Jefimenko's equations (or the
equivalent).  Our procedure avoids completely the use of
electromagnetic potentials. Specifically, we shall obtain the
electric dipole, the magnetic dipole, and the electric quadrupole
terms of the multipole expansion due to the radiation fields of an
arbitrary localized source.

\section{Jefimenko's equations from Maxwell's equations}

In this section we present three methods of calculating Jefimenko's
equations
directly from Maxwell's equations. The first method closely follows Ref.~\onlinecite{Jackson3nd}. The second method makes use of a Fourier
transformation as discussed by Ref.~\onlinecite{BornaticiBellotti1996}. For our purposes it suffices to
do a Fourier transformation only in the temporal coordinate.
We shall see that this method, although longer than the previous
one, avoids any possibility of misleading
manipulations with retarded quantities. The subtleties in
calculations involving retarded quantities have been discussed in Refs.~\onlinecite{Rohrlich2002A,Jefimenko2002,Rohrlich2002B}. We then present an
alternative method that is a variation of the first one; the main difference is the order of integration.

\subsection{\label{sec2a}Direct calculation using the retarded Green function}

The following approach is similar to that in
Refs.~\onlinecite{Jackson3nd}, \onlinecite{Heras95} and
\onlinecite{Rohrlich2002A}.  Maxwell's equations with sources in
vacuum are given by
\begin{align}\label{Gauss}
\grad \cdot{\bf E}&=4\pi\rho \\
\grad \cdot{\bf B}&=0\\
\label{Faraday}
 \grad \times{\bf E}&= -\frac{1}{c}\frac{\partial
{\bf B}}{\partial t} \\
\label{Ampere-Maxwell} \grad \times{\bf B}&= \frac{4\pi}{c}{\bf J}+
\frac{1}{c}\frac{\partial {\bf E}}{\partial t}.
\end{align}
Taking the curl of Eq.~\eqref{Faraday} and the time derivative of
Eq.~\eqref{Ampere-Maxwell}, we obtain
\begin{equation}\label{EquacaoParaE}
\Big(\grad^2-
\frac{1}{c^2}\frac{\partial^2}{\partial t^2}\Big) {\bf
E}=4\pi\Big(\grad \rho +
\frac{1}{c^2}\frac{\partial{\bf J}}{\partial t}\Big),
\end{equation}
where we used $\grad \times(\grad \times{\bf F}) = \grad (\grad
\cdot{\bf F}) - \grad ^2{\bf F}$, with $\grad ^2{\bf F}=
 (\grad ^2F_x)\hat{\bf x} + (\grad ^2F_y)\hat{\bf y} +
  (\grad ^2F_z)\hat{\bf z}$ and ${\bf F}$ being an arbitrary function.
  Similarly, we obtain for the magnetic field:
\begin{equation}\label{EquacaoParaB}
\Big(\grad ^2-
\frac{1}{c^2}\frac{\partial^2}{\partial t^2}\Big) {\bf
B}=-\frac{4\pi}{c}{\bf J}.
\end{equation}
The solutions of Eqs.~(\ref{EquacaoParaE}) and (\ref{EquacaoParaB})
can be obtained with the aid of the retarded Green function
$G_{ret}({\bf x},t;{\bf x}',t')$,  which satisfies the inhomogeneous
differential equation
\begin{equation}
\Big(\grad^2- \frac{1}{c^2}\frac{\partial^2}{\partial t^2}\Big)
G_{ret}({\bf x},t;{\bf x}',t')= \delta({\bf x}-{\bf
x}')\delta(t-t'),
\end{equation}
and is zero for $t-t'<0$. The solution for $G_{ret}$ for $t-t'> 0$
is given by\cite{Jackson2nd}
\begin{equation}\label{GR}
G_{ret}({\bf x},t;{\bf x}',t') =
-\frac{1}{4\pi}\frac{\delta\big(t'-(t-\vert{\bf x}-{\bf x}'\vert/c
\big))} {\vert{\bf x}-{\bf x}'\vert} =
-\frac{1}{4\pi}\frac{\delta\big(t'- (t-R/c)\big)}{R},
\end{equation}
where $R\equiv\vert{\bf R}\vert= \vert{\bf x}-{\bf x}'\vert$.
With the help of Eq.~(\ref{GR}) the
solution of Eq.~(\ref{EquacaoParaE}) can be written as
\begin{align}
{\bf E}({\bf x},t)&= -\!\int\frac{d{\bf x}'}{R} \!\int
dt' \delta\big(t'-(t-{R}/{c})\big)
\left(\grad'
\rho({\bf x}',t')
+ \frac{1}{c^2}\frac{\partial{\bf J({\bf
x}',t')}}{\partial t'} \right) \label{EField11} \\
&=
-\!\int\! d{\bf x}'
\frac{[\grad'\rho]} {R} -
\!\int d{\bf x}'\;\frac{[\dot{{\bf J}}]}{c^2\, R},\label{EField1}
\end{align}
where the notation $[\ldots]$ means that the quantity inside the
brackets is a function of ${\bf x}'$ and is evaluated
at the retarded time $t'=t-\vert{\bf x}-{\bf
x}'\vert/c$.

At this point much care must be taken, because
 $\grad'[\rho]\ne [\grad'\rho]$. Due to a bad choice of notation,
 the author in Ref.~\onlinecite{Rohrlich2002A} used, incorrectly,
 the quantity $[\grad'\rho]$ as if it were $\grad'[\rho]$ and,
 after an integration by parts, an incomplete
result for the electric field was found, as pointed out in
Ref.~\onlinecite{Jefimenko2002}. The correct relation is given by
$\grad'[\rho]= [\grad'\rho] + \hat{\bf R}[\dot \rho]/c$ (see, for
instance, Ref.~\onlinecite{Jackson3nd}). If we use the correct
relation, Eq.~(\ref{EField1}) becomes
\begin{align}
{\bf E}({\bf x},t) &= -\!\int\! d{\bf x}'
\frac{\grad'[\rho]}{R} +
\!\int d{\bf x}' \frac{[\dot\rho]\,\hat{\bf R}}{c R} -
\!\int d{\bf x}' \frac{[\dot{\bf J}]}{c^2\, R}\\
&=
\!\int d{\bf x}'\frac{[\rho] \hat{\bf R}}{R^2} +
\!\int d{\bf x}'\frac{[\dot\rho] \hat{\bf R}}{c R}-
\!\int d{\bf x}'\frac{[\dot{{\bf J}}]}{c^2 R}. \label{ERetardado2}
\end{align}
In the last step we integrated by parts and
discarded surface terms, because the charge distribution is localized.
 Equation~\eqref{ERetardado2} is one of the Jefimenko's equations.
\cite{JefimenkoBook,Jackson3nd} Note the retarded character of the
electric field. The first term on the right-hand side of Eq.~\eqref{ERetardado2} is
the retarded Coulomb term.

As shown by Panofsky and Phillips,
\cite{PanofskyPhillips} there is an equivalent way of deriving Eq.~\eqref{ERetardado2} for the electric field which manifestly shows
the transverse character of the radiation field. A
discussion of this point can be found in Ref.~\onlinecite{McDonald}; we will
discuss this point in Sec.~III.

To obtain the desired expression for the magnetic field, we use the
retarded Green function in Eq.~(\ref{GR}) to rewrite
Eq.~(\ref{EquacaoParaB}) as
\begin{equation}
\label{BRetardado1}
{\bf B}({\bf x},t) = \frac{1}{c}\!\int\frac{d{\bf x}'}{R}
\!\int dt' \delta\big(t'-(t-{R}/{c})\big)
\grad'\times {\bf J}({\bf
x}',t')
=\frac{1}{c}\!\int\frac{d{\bf x}'}{R}
[\grad'\times {\bf J}].
\end{equation}
If we use the relation
$\grad'\times [{\bf J}] =
[\grad'\times {\bf J}] +
\dfrac{\hat{\bf R}}{c}\times[\dot{\bf J}]$
(see Ref.~\onlinecite{Jackson3nd}), Eq.~(\ref{BRetardado1})
takes the form
\begin{equation}\label{BRetardado2}
{\bf B}({\bf x},t)= \frac{1}{c}\!\int\frac{d{\bf x}'}{R}
\grad'\times [{\bf J}] -
\frac{1}{c^2}\!\int d{\bf x}' \frac{\hat{\bf
R}\times[\dot{\bf J}]}{R}.
\end{equation}
We integrate by parts and obtain
\begin{equation}\label{BRetardado3}
{\bf B}({\bf x},t)= \frac{1}{c}\!\int d{\bf x}'\,
\grad'\times
\Big(\frac{[{\bf J}]}{R} \Big) + \frac{1}{c}\!\int d{\bf
x}'\, [{\bf J}]\times
\grad'\Big(\frac{1}{R} \Big)
- \frac{1}{c^2}\!\int d{\bf x}' \frac{\hat{\bf
R}\times[\dot{\bf J}]}{R}.
\end{equation}
The first term on the right-hand side of Eq.~(\ref{BRetardado3}) vanishes
because the current distribution is localized in space. We use the relation
$\grad'(1/R) =
-\hat{\bf R}/R^2$ to obtain
\begin{equation}\label{FinalBField}
{\bf B}({\bf x},t)= \!\int d{\bf x}'\left[ \frac{[{\bf
J}]\times\hat{\bf R}} {c\,R^2} + \frac{[\dot{\bf J}]\times\hat{\bf
R}} {c^2\,R} \right].
\end{equation}
The first term on the right-hand side of Eq.~\eqref{BRetardado3} is
the retarded Biot-Savart term; the transverse radiation field is
contained in the last term. Equations~(\ref{ERetardado2}) and
(\ref{FinalBField}) are the Jefimenko equations.

\subsection{Fourier method}

In Sec.~\ref{sec2a} we showed how to obtain Jefimenko's
equations by a careful
treatment of the derivatives of retarded
quantities. This point is crucial --
spatial derivatives cannot be commuted with retarding the functions, because the
retarded function depends on the coordinates in its time argument.
A simple way to circumvent this difficulty is to use Fourier
transforms and factor out the time dependence of the
functions so these subtleties are not encountered. We will show how
to obtain the electric field and will leave the derivation of the
magnetic field as an exercise for the interested reader (see,
for example, Ref.~\onlinecite{BornaticiBellotti1996}).

We start with the electric field given by Eq.~(\ref{EField11}). The
term involving the current density takes the form after integration
over time,
\begin{equation}\label{EField2}
-\!\int\frac{d{\bf x}'}{R} \!\int dt'
\delta\Big(t'-(t-{R}/{c})\Big)
\frac{1}{c^2}\frac{\partial{\bf J}}
{\partial t'}({\bf x}',t')
=-\!\int d{\bf x}' \frac{[\dot{{\bf J}}]}{c^2\, R}.
\end{equation}
We introduce the Fourier transformation $\tilde\rho({\bf
x}',\omega)$  of $\rho({\bf x}',t')$ as $ \rho({\bf
x}',t')=\!\int\!\tilde\rho({\bf x}',\omega) e^{-i\,\omega t'}d\omega
$, and express the remaining contribution to the electric field in
Eq.~(\ref{EField1}) as
\begin{subequations}
\label{Fourier1}
\begin{align}
-\!\int\frac{d{\bf x}'}{R}&\!\int\!
dt'
\delta\big(t'-(t-R/c)\Big)
\grad' \rho({\bf
x}',t') \nonumber \\
&=-\!\int\frac{d{\bf x}'}{R} \!\int dt' \delta\big(t'-(t-R/c)\big)
\!\int d\omega\grad' \tilde\rho({\bf
x}',\omega) e^{-i \omega t'} \label{line1}\\
&= -\!\int d\omega e^{-i\omega t}\!\int{d{\bf x}'} \frac{e^{i k
R}}{R} \grad' \tilde\rho({\bf
x}',\omega) \label{line2}\\
&= -\!\int d\omega e^{-i \omega t}\!\int{d{\bf x}'} \Bigg\{\grad'
\Big(\frac{e^{i k R}}{R}\tilde\rho({\bf x}',\omega) \Big) -
\tilde\rho({\bf x}',\omega) \Big(\frac{\hat{\bf
R}}{R^2}- ik\frac{\hat{\bf R}}{R} \Big)e^{i k R} \Bigg\}\label{line3} \\
&= \! \int d{\bf x}'\frac{\hat{\bf R}}{R^2} \!\int d\omega
\tilde\rho({\bf x}',\omega) e^{-i\omega(t-R/c)}+ \!\int d{\bf
x}'\frac{\hat{\bf R}}{c R}\!\int d\omega \tilde\rho({\bf
x}',\omega) (-i\omega) e^{-i\omega(t-R/c)}
\label{line4}\\
&= \!\int d{\bf x}'\frac{\hat{\bf R}[\rho]}{R^2} + \!\int d{\bf
x}'\frac{\hat{\bf R}[\dot\rho]}{c R},
\end{align}
\end{subequations}
To obtain Eq.~\eqref{line3}
we integrated by parts; to obtain Eq.~\eqref{line4} we
discarded the surface term.

We combine Eq.~(\ref{EField11}) with Eqs.~(\ref{EField2}) and
(\ref{Fourier1}) and obtain the electric field generated by
arbitrary (but localized) sources, namely, Eq.~(\ref{ERetardado2}).
On the right-hand side of Eq.~(\ref{line1}) it is obvious that
$\grad'$ acts only on $\tilde\rho({\bf x}',\omega)$. Hence, in
Eq.~\eqref{line3} there is no possibility of thinking that $\grad'$
also acts on the exponential $e^{ikR}$., The subtleties of dealing
with retarded quantities are circumvented by the Fourier method.

\subsection{Postponing the delta-function integration}

When we are faced with integrals involving delta functions, we are
usually tempted to do them first, because they are so easy. To
derive Jefimenko's equation for the electric field, given by
Eq.~\eqref{ERetardado2}, this procedure is not optimum. Let us start
with Eq.~(\ref{EField11})
\begin{equation}\label{Delta1}
{\bf E}({\bf x},t)= -\!\int\frac{d{\bf x}'}{R}\!\int
dt' \delta\big(t'-(t-{R}/{c})\big)
\Big(\grad'\rho({\bf
x}',t') + \frac{1}{c^2}\frac{\partial{\bf J}}
{\partial t'}({\bf x}',t') \Big).
\end{equation}
Instead of first performing the time integration using the Dirac
delta function, we integrate by parts on both terms on
the right-hand side of Eq.~(\ref{Delta1}) so that it takes the form
\begin{align}
{\bf E}({\bf x},t)&= \!\int\! d{\bf x}'\!\int\!
dt'\Bigg\{
\delta'\big(t'-(t-{R}/{c})\big)
\Big(\frac{1}{Rc} \rho({\bf x}',t')
\grad' {R}
+ \frac{1}{Rc^2}{\bf J}({\bf x}',t')\Big) \nonumber \\
& \quad +
\delta\big(t'-(t-{R}/{c})\big)
\Big(\grad' \frac{1}{R} \Big)
\rho({\bf x}',t')\Bigg\}.
\end{align}
If we use the relations
$\grad' {R} =
-\hat{\bf R}$
and
$\grad' 1/R =
\hat{\bf R}/R^2$, we obtain
\begin{align}
{\bf E}({\bf x},t)&= \!\int d{\bf x}' \!\int
dt'\Bigg\{
\delta'\big(t'-(t-{R}/{c})\big)
\left(- \frac{\hat{\bf R}}{Rc}
\rho({\bf x}',t') + \frac{1}{Rc^2} {\bf J}({\bf
x}',t')\right) \nonumber \\
& \quad +
\delta\Big(t'-(t-{R}/{c})\Big)
\rho({\bf x}',t') \frac{\hat{\bf R}}{R^2} \Bigg\}.
\end{align}
By using the properties of the Dirac delta function, we can easily
perform the integration over $t'$ to obtain
Eq.~(\ref{ERetardado2}). An analogous calculation provides an expression for the magnetic field.

\section{Multipole radiation via Jefimenko's equations}

The main purpose of this section is to add to the list of problems
that can be handled directly with Jefimenko's equations by
calculating the first multipole contributions to the radiation
fields of an arbitrary localized source. We shall obtain the first
three contributions, namely, the electric dipole, the magnetic
dipole and the electric quadrupole terms. Most textbooks treat this
problem by calculating first the electromagnetic potentials.

Although Eq.~(\ref{ERetardado2}) for the electric field gives the
correct expression for the electric field of moving charges, it is
preferable for our purposes to write it in an equivalent form as
given in Refs.~\onlinecite{PanofskyPhillips} and
\onlinecite{McDonald}:
\begin{equation}
\label{CampoEMcDonald}
\textbf{E}(\textbf{x},t) = \!\int d
\textbf{x}'\frac{[\rho]\,\hat{\textbf{R}}}{R^2} + \!\!\int d
\textbf{x}'\frac{([\textbf{J}]\cdot\hat{\textbf{R}})\hat{\textbf{R}}
+([\textbf{J}]\times\hat{\textbf{R}})\times\hat{\textbf{R}}
}{cR^{2}} + {\int d
\xv'\frac{(\dot{[\textbf{J}]}\times\hat{\textbf{R}})\times\hat{\textbf{R}}
}{c^{2}R}}.
\end{equation}
We obtain Eq.~\eqref{CampoEMcDonald}
starting with Jefimenko's equation for the electric field
in Eq.~(\ref{ERetardado2}). Our derivation will follow the one in
Ref.~\onlinecite{McDonald}. Note that (the Einstein convention of implicit
summation over repeated indices is assumed)
\begin{subequations}
\begin{align}
\grad'\cdot[{\bf J}]
&=\partial'_i J_i(\xv', t')
\Big|_{t' = t - R/c} +
\frac{\partial}{\partial t'}
J_i(\xv', t')
\Big|_{t' = t - R/c}
\Big( -\frac{1}{c} \partial'_i R\Big)\\
&=
[\grad'
\cdot{\bf J}] + [\dot J_i] \Big(\frac{X_i}{cR}\Big)=
-[\dot\rho] + [\dot{\bf J}]\cdot\frac{\hat{\bf R}}{c}, \label{eq:above}
\end{align}
\end{subequations}
where $X_i\equiv x_i - x'_i$ and in the last step we used the
continuity equation. From Eq.~\eqref{eq:above} we have
$[\dot\rho] = - \grad'
\cdot[{\bf J}] +
[\dot{\bf J}]\cdot{\hat{\bf R}}/{c}$, which can be substituted into
the second term on the right-hand side of Eq.~(\ref{ERetardado2}) to
yield
\begin{equation}\label{Extra2}
\frac{1}{c}\!\int\! \frac{[\dot\rho]\, \hat{\bf R}}{R} d\xv' = \frac{1}{c}\!\int\! \frac{\Big(\grad'
\!\cdot[{\bf J}]\Big)\, \hat{\bf R}}{R} d\xv' +
\frac{1}{c^2}\!\int\!
\frac{\Big([\dot{\bf J}]\cdot\hat{\bf R}\Big)\hat{\bf R}}{R} d{\bf x}'.
\end{equation}
We show that the first term on the right-hand side of Eq.~\eqref{Extra2} is proportional to $1/R^2$, so that it does not contribute to
the radiation field. Observe that
\begin{subequations}
\label{Extra3}
\begin{align}
- \frac{1}{c}\!\int
\frac{\Big(\grad'
\!\cdot[{\bf J}]\Big) \hat{\bf R}}{R} d{\bf x}'
&=
-\frac{\hat{\bf e}_k}{c}\!\int\Big(\partial '_i
[J_i]\Big)\frac{X_k}{R^2} d{\bf x}' \\
&=
-\frac{\hat{\bf e}_k}{c} \!\int\partial '_i
\left([J_i]\frac{X_k}{R^2}\right) d{\bf x}' +
\frac{\hat{\bf e}_k}{c} \!\int [J_i]\partial '_i
\left(\frac{X_k}{R^2}\right) d{\bf x}' \label{Extra3.second}\\
&=
0 + \frac{\hat{\bf e}_k}{c} \!\int [J_i]
\left\{X_k\,\frac{2}{R^3}\,\frac{X_i}{R} -
\frac{\delta_{ki}}{R^2}\right\}d{\bf x}' \\
&=
\frac{1}{c}\!\int\frac{2\Big([{\bf J}]\cdot{\hat{\bf R}}\Big)
\hat{\bf R} - [{\bf J}]}{R^{\, 2}} d{\bf x}' \\
&=
\frac{1}{c}\!\int\frac{\Big([{\bf J}]\cdot{\hat{\bf R}}\Big)
\hat{\bf R} + \Big([{\bf J}]\times\hat{\bf R}\Big)
\times\hat{\bf R}}
{R^2} d{\bf x}',
\end{align}
\end{subequations}
where we have used the identity $({\bf a}\times{\bf b})\times{\bf c}
= ({\bf a}\cdot {\bf c})\,{\bf b} - ({\bf b}\cdot{\bf c})\,{\bf a}$,
and the fact that the surface term appearing in
Eq.~\eqref{Extra3.second} vanishes because the sources are
localized. We substitute Eq.~(\ref{Extra3}) into Eq.~(\ref{Extra2})
and insert the result into Eq.~(\ref{ERetardado2}) to obtain
\begin{subequations}
\begin{align}
\textbf{E}(\textbf{x},t)
&=
\!\int \frac{[\rho]\,\hat{\textbf{R}}}{R^2}\,d\textbf{x}' +
\frac{1}{c}\!\int \frac{([\textbf{J}]\cdot\hat{\textbf{R}})\,\hat{\textbf{R}}
+([\textbf{J}]\times\hat{\textbf{R}})\times\hat{\textbf{R}}
}{R^{2}}\,d\textbf{x}' \nonumber \\
&\quad +
\frac{1}{c^2}\!\int
\frac{\big([\dot{\bf J}]\cdot\hat{\bf R}\big)\hat{\bf R}}{R}
\, d{\bf x}' -
\frac{1}{c^2}\!\int
\frac{[\dot{{\bf J}}]}{R} d{\bf x}^{\prime}\\
&=
\!\int d
\textbf{x}'\frac{[\rho]\,\hat{\textbf{R}}}{R^2} + \!\int d
\textbf{x}'\frac{([\textbf{J}]\cdot\hat{\textbf{R}})\hat{\textbf{R}}
+([\textbf{J}]\times\hat{\textbf{R}})\times\hat{\textbf{R}}
}{cR^{2}} + \!\int d
\textbf{x}'\frac{(\dot{[\textbf{J}]}\times\hat{\textbf{R}})\times\hat{\textbf{R}}
}{c^{2}R},
\end{align}
\end{subequations}
which is Eq.~(\ref{CampoEMcDonald}). If we consider arbitrary
time-varying sources at rest and use Eqs.~(\ref{FinalBField}) and
(\ref{CampoEMcDonald}), the (transverse) magnetic and electric
radiation fields are given by
\begin{equation}
\textbf{B}_{\rm rad}({\bf x},t) =
\frac{1}{c}\!\int d\textbf{x}'
\frac{[\dot{\textbf{J}}]\times\hat{\textbf{R}}}{Rc},\label{radb}
\quad \textbf{E}_{\rm rad}({\bf x},t) =
{\int d
\textbf{x}'\frac{(\dot{[\textbf{J}]}\times\hat{\textbf{R}})\times\hat{\textbf{R}}
}{c^{2}R}}.
\end{equation}
It can be shown\cite{Heras1998} that for time-varying sources in
motion, Eq.~(\ref{radb}) gives the radiation fields plus additional
non-radiative terms of order ${\cal O}(1/R^2)$. For instance, if a
Hertz dipole is accelerated, integration of equations
Eq.~(\ref{radb}) yields radiation fields plus nonradiative terms of
order ${\cal O}(1/R^2)$ which are induced by the dipole motion (for
details, see Refs.~\onlinecite{Heras1998}).

In the radiation zone we can write $\hat{\bf R}\simeq\hat{\bf x}$,
${1}/{R}\simeq{1}/{r}$ and $R\simeq r-\hat{\bf x}\cdot{\bf x}'$,
 where we defined $r=\vert{\bf x}\vert$.
 If we substitute these approximations into Eq.~\eqref{radb}, we
obtain
\begin{align}\label{Brad}
{\bf B}_{\rm rad}({\bf x},t) &\simeq \frac{1}{c^{2}r}\!\int d{\bf
x}' \,\dot{{\bf J}}\Big({\bf x}',t_{0}+\frac{\hat{\bf x} \cdot{\bf
x}'}{c}\Big)\times\hat{{\bf x}}
\\
{\bf E}_{\rm rad}({\bf x},t) &\simeq \frac{1}{c^{2}r}\!\int d{\bf
x}' \Big[\dot{{\bf J}}\Big({\bf x}',t_{0}+\frac{\hat{\bf x}
\cdot{\bf x}'}{c}\Big)\times\hat{{\bf x}}\Big] \times\hat{{\bf
x}},\label{Brad2}
\end{align}
where $t_{0}=t - {r}/{c}$ is the retarded time of the origin.
A comparison of Eqs.~\eqref{Brad} and \eqref{Brad2} leads to the relation
\begin{equation}\label{EBr}
\textbf{E}_{\rm rad}({\bf x},t) = \textbf{B}_{\rm rad}({\bf x},t)
\times\hat{\textbf{x}}.
\end{equation}

Now we are ready to calculate the first multipole contributions for
the radiation fields. We need to
calculate only one of the radiation fields because the other
is readily obtained by Eq.~(\ref{EBr}), which
also shows that the fields in the radiation zone
are mutually orthogonal. We start by calculating
the electric dipole term and then consider the next order
contribution given by both the magnetic dipole and electric
quadrupole terms.

\subsection {The electric dipole contribution}

The lowest order contribution to the radiation fields comes from the
electric dipole term. For simplicity, we calculate the lowest order
contribution to the radiation magnetic field, which we denote by
$\textbf{B}_{\rm rad}^{(1)}$,
\begin{equation}\label{Brad1}
\textbf{B}_{\rm rad}^{(1)}({\bf x},t) = \frac{1}{c^{2}r}\left\{\int
d \textbf{x}' \dot{\textbf{J}}(\textbf{x}',
t_{0})\right\}\times\hat{\textbf{x}}.
\end{equation}
We write the unit vectors of the cartesian basis as ${\hat{\bf e}}_i
= \grad'x_{i}'$, $(i=1,2,3)$, and write any vector ${\bf v}$ as
${\bf v} = \hat{\bf e}_i v_i = \hat{\bf e}_i({\bf v}\cdot\hat{\bf
e}_i)$. The integral in Eq.~\eqref{Brad1} can be expressed as
\begin{subequations}
\begin{align}\label{Integral1}
\int d \textbf{x}' \dot{\textbf{J}}(\textbf{x}', t_{0}) &=
\hat{\textbf{e}_{i}} \!\int d
\textbf{x}'\,\dot{\textbf{J}}(\textbf{x}',
t_{0})\cdot\hat{\textbf{e}_{i}} = \hat{\textbf{e}_{i}} \!\int d
\textbf{x}'\,\dot{\textbf{J}}(\textbf{x}', t_{0})\cdot\nabla'x_{i}'
\\
&= \hat{\textbf{e}_{i}} {\int d \textbf{x}'
\grad'\cdot\Big(x_{i}'\,\dot{\textbf{J}}(\textbf{x}', t_{0}) \Big) -
\hat{\textbf{e}_{i}}} \!\int d \textbf{x}'\,
x_{i}'\,\nabla'\cdot\dot{\textbf{J}}(\textbf{x}', t_{0}),
\end{align}
\end{subequations}
where in the last step we integrated by parts. Because we are
considering localized sources, the first integral
on the right-hand side of Eq.~(\ref{Integral1}) vanishes (after the use of Gauss'
theorem this integral is converted to a zero surface term). The remaining integral
may be cast into a convenient form if we use the relation
\begin{equation}\label{ConseqCE}
\grad'\cdot\dot{\textbf{J}}(\textbf{x}',
t_{0})=-\frac{\partial^{2}\rho(\textbf{x}', t_{0})}{\partial t^2},
\end{equation}
which is a direct consequence of the continuity equation. To obtain
Eq.~(\ref{ConseqCE}) we used the relation
$\partial\rho({\bf x}',t_0)/\partial t =
\partial\rho({\bf x}',t_0)/\partial t_0$ (see Ref.~\onlinecite{Heras2007}, note 5). We substitute Eqs.~(\ref{ConseqCE}) and
(\ref{Integral1}) into Eq.~(\ref{Brad1}) and obtain
\begin{equation}\label{BdipoloEletrico}
\textbf{B}_{\rm rad}^{(1)}({\bf x},t) =
\frac{1}{c^{2}r}\frac{\partial^{2}}{\partial
t^2}\overbrace{\left\{\int d \textbf{x}' \rho(\textbf{x}',
t_{0})\,{\bf x}^{\,\prime}\right\}}^{\textbf{p}(t_{0})}
\times\,\hat{\textbf{x}}=\frac{\ddot{\textbf{p}}(t_{0})
\times\hat{\textbf{x}}}{c^{2}r},
\end{equation}
where ${\bf p}(t_0)$ is the electric dipole moment of the
distribution at the retarded time $t_0$.
Now it is clear why this first term is called the electric dipole term.
The radiation electric field is readily obtained from Eq.~(\ref{EBr})
\begin{equation}\label{EdipoloEletrico}
\textbf{E}_{\rm rad}^{(1)}({\bf x},t) =
\frac{\left[\ddot{\textbf{p}}(t_{0})\times\hat{\textbf{x}}\right]
\times\hat{\textbf{x}} }{c^2r}.
\end{equation}

Equations~(\ref{Brad1}) and (\ref{EdipoloEletrico}) are
the radiation fields of the electric dipole term, that is, the first-order contribution to the multipole expansion. These
expressions are valid for arbitrary but localized,
sources in vacuum such as an oscillating
electric dipole.

\subsection{Next order contribution}

To calculate the next order term we need to take into account the
second term of the expansion
\begin{equation}\label{TermoOrdem2}
\dot{{\bf J}}\Big({\bf x}',t_{0} + \frac{\hat{\bf x}\cdot{\bf
x}'}{c} \Big) \approx \dot{{\bf J}}({\bf x}',t_{0}) + \frac{\hat{\bf
x} \cdot{\bf x}'}{c} \,\ddot{{\bf J}}({\bf x}',t_{0}).
\end{equation}

We substitute Eq.~\eqref{TermoOrdem2} into Eq.~(\ref{Brad}) and
identify the next order contribution to the radiation magnetic
field, ${\bf B}^{(2)}_{\rm rad}$, given by
\begin{equation}\label{Brad2}
\textbf{B}_{\rm rad}^{(2)}({\bf x},t) = \frac{1}{c^{3}r}\left(\int d
\textbf{x}'\,\ddot{\textbf{J}}(\textbf{x}', t_{0})
(\hat{\textbf{x}}\cdot\textbf{x}')\right) \times\hat{\textbf{x}}.
\end{equation}
For reasons that will become clear we will split the integral in
Eq.~\eqref{Brad2} into antisymmetric and symmetric contributions
under the exchange of \textbf{J} and $\textbf{x}'$. This
rearrangement will give rise to the magnetic dipole and electric
quadrupole terms of the multipole expansion for the radiation
fields.

If we write $\ddot{\textbf{J}}(\textbf{x}',
t_{0})\,(\hat{\textbf{x}}\cdot\textbf{x}')$ as
$\dfrac{1}{2}\,\ddot{\textbf{J}}(\textbf{x}',
t_{0})\,(\hat{\textbf{x}}\cdot\textbf{x}') + \dfrac{1}{2}
\ddot{\textbf{J}}(\textbf{x}', t_{0})
(\hat{\textbf{x}}\cdot\textbf{x}')$ and sum and subtract
$\dfrac{1}{2}[\ddot{\textbf{J}}(\textbf{x}',
t_{0})\cdot\hat{\textbf{x}}]\,\textbf{x}'$ in the integrand of the
right-hand side of Eq.~(\ref{Brad2}), we obtain
\begin{eqnarray}\label{BradSimEAnti}
\textbf{B}_{\rm rad}^{(2)}({\bf x},t) &=&
\frac{1}{c^{2}r}\left\{\frac{1}{2c}\!\int d
\textbf{x}'\left[\ddot{\textbf{J}}(\textbf{x}',
t_{0})(\hat{\textbf{x}}\cdot\textbf{x}') -
\Big(\ddot{\textbf{J}}(\textbf{x}',
t_{0})\cdot\hat{\textbf{x}}\Big)\textbf{x}'\right]\right\}
\times\hat{\textbf{x}} \cr\cr &&{}+
\frac{1}{c^{2}r}\left\{\frac{1}{2c}\!\int d
\textbf{x}'\left[\ddot{\textbf{J}}(\textbf{x}',
t_{0})(\hat{\textbf{x}}\cdot\textbf{x}') +
\Big(\ddot{\textbf{J}}(\textbf{x}',
t_{0})\cdot\hat{\textbf{x}}\Big)\textbf{x}'\right]\right\}\times\hat{\textbf{x}}.
\end{eqnarray}
For pedagogical reasons we shall treat the magnetic
dipole and electric quadrupole cases separately.

Consider the antisymmetric term of
Eq.~(\ref{BradSimEAnti}). We denote this contribution by ${\bf
B}^{(2a)}_{\rm rad}$, use
$({\bf a}\times{\bf b})\times{\bf c} = ({\bf
c}\cdot{\bf a}){\bf b}
- ({\bf c}\cdot{\bf b}){\bf a}$, and write it in the following
suggestive way
\begin{subequations}
\begin{align}
\textbf{B}_{\rm rad}^{(2a)}({\bf x},t) &=
\frac{1}{c^{2}r}\frac{1}{2c}\!\int d
\textbf{x}'\left[\ddot{\textbf{J}}(\textbf{x}',
t_{0})\hat{\textbf{x}}\cdot\textbf{x}' -
\Big(\ddot{\textbf{J}}(\textbf{x}',
t_{0})\cdot\hat{\textbf{x}}\Big)\textbf{x}'\right]\times\hat{\textbf{x}} \\
&= \frac{1}{c^{2}r} \Bigg\{\;\underbrace{\left(\int d
\textbf{x}'\frac{\textbf{x}'\times\ddot{\textbf{J}}(\textbf{x}',
t_{0})}{2c}\right)}_{\ddot{\textbf{m}}(t_0)}
\times\,\hat{\textbf{x}} \Bigg\}\times\hat{\textbf{x}},
\end{align}
\end{subequations}
where we have identified the second time derivative of the magnetic dipole
moment of the charge distribution at the retarded
time as $\ddot{\bf m}(t_0)$. Hence, the antisymmetric contribution
for the radiation magnetic field is given by
\begin{equation}\label{Brad2AntiFinal}
\textbf{B}_{\rm rad}^{(2a)}({\bf x},t) = \frac{[\ddot{\bf
m}(t_0)\times\hat{\bf x}]\times\hat{\bf x}} {c^2 r}.
\end{equation}
The appearance of the magnetic dipole moment of the distribution
justifies the name given to this contribution. The corresponding
radiation electric field is readily given by
\begin{equation}\label{Erad2AntiFinal}
\textbf{E}_{\rm rad}^{(2a)}({\bf x},t) = \frac{\hat{\bf
x}\times\ddot{\textbf{m}}(t_0)}{c^{2}r}
\end{equation}
Expressions (\ref{Brad2AntiFinal}) and (\ref{Erad2AntiFinal}) are
valid for an arbitrary, but localized, time-varying source at rest
in vacuum such as an oscillating magnetic
dipole.

We denote the symmetric term of
Eq.~(\ref{BradSimEAnti}) as ${\bf
B}^{(2s)}_{\rm rad}$. We have
\begin{equation}
\textbf{B}_{\rm rad}^{(2s)}({\bf x},t) =
\frac{1}{c^{2}r}\frac{1}{2c}\!\int d
\textbf{x}'\left\{\ddot{\textbf{J}}(\textbf{x}',
t_{0})\,(\hat{\textbf{x}}\cdot\textbf{x}') +
\Big(\ddot{\textbf{J}}(\textbf{x}', t_{0})\cdot\hat{\textbf{x}}\Big)
\textbf{x}'\right\} \times\hat{\textbf{x}}.\label{qe}
\end{equation}
We manipulate the first integral on the right-hand side
of Eq.~\eqref{qe} in the same way as before. We have
\begin{subequations}
\begin{align}
\int \! d \textbf{x}'\ddot{\textbf{J}}(\textbf{x}',
t_{0})\,(\hat{\textbf{x}}\cdot\textbf{x}') &= \hat{\textbf{e}_{i}}
\!\int \! d \textbf{x}' \,(\hat{\textbf{x}}\cdot\textbf{x}')
\ddot{\textbf{J}}(\textbf{x}', t_{0})
\cdot\hat{\textbf{e}_{i}}\\
&= \hat{\textbf{e}_{i}} \!\int \! d
\textbf{x}'\,(\hat{\textbf{x}}\cdot\textbf{x}')
\ddot{\textbf{J}}(\textbf{x}', t_{0})\cdot\nabla'x_{i}'.
\end{align}
\end{subequations}
If we integrate by parts and remember that the surface term
vanishes, we obtain
\begin{subequations}
\begin{align}
\int d \textbf{x}'\ddot{\textbf{J}}(\textbf{x}',
t_{0})(\hat{\textbf{x}}\cdot\textbf{x}') &= -\hat{\textbf{e}_{i}}
\!\int\! d \textbf{x}' \,
x_{i}'\grad'\cdot\Big[(\hat{\textbf{x}}\cdot\textbf{x}')
\ddot{\textbf{J}}(\textbf{x}', t_{0})\Big] \\
&= -\hat{\textbf{e}_{i}} \!\int\! d \textbf{x}' \,
x_{i}'\left[\Big(\nabla'(\hat{\textbf{x}}\cdot\textbf{x}')\Big)
\cdot\ddot{\textbf{J}}(\textbf{x}',
t_{0})+(\hat{\textbf{x}}\cdot\textbf{x}')\,\nabla'\cdot
\ddot{\textbf{J}}(\textbf{x}', t_{0})\right].
\end{align}
\end{subequations}
We use $ \grad'\cdot \ddot{\textbf{J}}(\textbf{x}', t_{0}) =
-\partial^{3}\rho({\bf x}',t_0)/\partial t^{3}$, the
relation\cite{Heras2007} $\partial\rho({\bf x}',t_0)/\partial t =
\partial\rho({\bf x}',t_0)/\partial t_0$), and
$\nabla'(\hat{\textbf{x}}\cdot\textbf{x}') = \hat{\textbf{x}}$ to
obtain
\begin{equation}
\int\! d \textbf{x}'\,\ddot{\textbf{J}}(\textbf{x}',
t_{0})(\hat{\textbf{x}}\cdot\textbf{x}') =
-\hat{\textbf{e}_{i}}\!\int\! d \textbf{x}'\,\hat{\textbf{x}}
\cdot\ddot{\textbf{J}}(\textbf{x}',
t_{0})x_{i}'+\hat{\textbf{e}_{i}}\frac{\partial^{3}}{\partial t^{3}}
\int\! d \textbf{x}'\,(\hat{\textbf{x}}\cdot\textbf{x}')
x_{i}'\rho(\textbf{x}', t_{0}),\label{eq41}
\end{equation}
which implies
\begin{equation}
\int\! d \textbf{x}'\left[\ddot{\textbf{J}}(\textbf{x}',
t_{0})(\hat{\textbf{x}}\cdot\textbf{x}') +
\Big(\ddot{\textbf{J}}(\textbf{x}',
t_{0})\cdot\hat{\textbf{x}}\Big)\textbf{x}'\right] =
\frac{\partial^{3}}{\partial t^{3}}\!\int\! d
\textbf{x}'(\hat{\textbf{x}}\cdot\textbf{x}')\textbf{x}'
\rho(\textbf{x}', t_{0}).
\end{equation}
The left-hand side of Eq.~(\ref{eq41}) is the
integral in Eq.~(\ref{qe}). We write
the factor $1/2$ in Eq.~(\ref{qe}) as $3/6$ and obtain
\begin{subequations}
\begin{align}\label{Brad2intermed}
\textbf{B}_{\rm rad}^{(2s)}({\bf x},t) &= \frac{1}{6c^{3}r}
\left\{\frac{\partial^{3}}{\partial t^{3}}\!\int\! d
\textbf{x}'(3\hat{\textbf{x}}\cdot\textbf{x}')
\textbf{x}'\rho(\textbf{x}', t_{0})\right\}
\times\hat{\textbf{x}}\\
&= \frac{1}{6c^{3}r} \Bigg\{\frac{\partial^{3}}{\partial t^{3}}
\int\! d \textbf{x}'(3(\hat{\textbf{x}}\cdot\textbf{x}')
\textbf{x}'+r'^{2}\hat{\textbf{x}})\rho(\textbf{x}', t_{0})
\Bigg\}\times\hat{\textbf{x}},
\end{align}
\end{subequations}
where in the last step we included in the integrand the term
$r'^{\,2}\,\hat{\textbf{x}}\,\rho(\textbf{x}', t_{0})$, where
$r^{\,\prime}=\vert{\bf r}^{\,\prime}\vert$, which gives a vanishing
contribution to the result because
$\hat{\textbf{x}}\times\hat{\textbf{x}}={\bf 0}$.
Now, we define the transformation ${\bf Q}$ such that
\begin{equation}\label{QuadrupoleOperator}
{\bf Q}(\mbox{{\mathversion{bold}${\xi}$}},t) = \!\int\! d
\textbf{x}' \left[3(\mbox{{\mathversion{bold}${\xi}$}}
\cdot\textbf{x}') \textbf{x}'+r'^{2}
\mbox{{\mathversion{bold}${\xi}$}} \right]\rho(\textbf{x}', t).
\end{equation}
This transformation takes a vector and
an instant of time $(\xib,t)$, and
transforms it into the vector ${\bf
Q}(\xib,t)$, given by Eq.~\eqref{QuadrupoleOperator}.
If we use the definition (\ref{QuadrupoleOperator}),
Eq.~(\ref{Brad2intermed}) for the antymmetric contribution for ${\bf
B}^{(2)}_{\rm rad}$ becomes
\begin{equation}\label{BQuadrupoloFinal}
\textbf{B}_{\rm rad}^{(2s)}({\bf x},t) = \frac{1}{6c^{3}r}
\stackrel{...}{\bf Q}(\hat{\bf x},t_0) \times\hat{\textbf{x}}.
\end{equation}
The corresponding symmetric contribution for the radiating electric
field is obtained from Eq.~(\ref{EBr}):
\begin{equation}\label{EQuadrupoloFinal}
\textbf{E}_{\rm rad}^{(2s)}({\bf x},t) =\frac{1}{6c^{3}r}
\left[\stackrel{...}{\bf Q}(\hat{\bf x},t_0)
\times\hat{\textbf{x}}\right] \times\hat{\bf x}.
\end{equation}
In order to interpret the above term, let us define a linear
operator ${\bf Q}^{t}$,  for a fixed instant $t$, by ${\bf
Q}^{t}(\xib) \equiv {\bf Q}(\xib,t)$. Note that ${\bf Q}^{t}$ is a
linear operator so that $ {\bf Q}^{t}(\alpha_1\xib_1 +
\alpha_2\xib_2) = \alpha_1{\bf Q}^{t}(\xib_1) + \alpha_2{\bf
Q}^{t}(\xib_2) $, for $\alpha_1$, $\alpha_2\in \cal R$. The linear
operator ${\bf Q}^{ t}$ is called the electric quadrupole operator.
Since $ {\bf Q}^{t}(\hat{\bf e}_i)$ is a vector in ${\cal R}^3$, we
can write it as a linear combination of the basis vectors, namely
\begin{equation}
{\bf Q}^{t}({\hat{\bf e}}_i) = \sum_{j=1}^3 {\bf
Q}^{t}_{ji}\,{\hat{\bf e}}_j \quad (i=1,2,3).
\end{equation}
The coefficients ${\bf Q}^{t}_{ji}$ are the cartesian elements of
the electric quadrupole tensor (a second rank tensor) of the
distribution at instant $t$. That is why we interpret
Eqs.~(\ref{BQuadrupoloFinal}) and (\ref{EQuadrupoloFinal}) as the
electric quadrupole contribution for the radiation fields.

The electric quadrupole radiation fields given by
Eqs.~(\ref{BQuadrupoloFinal}) and (\ref{EQuadrupoloFinal}), together
with the magnetic dipole radiation fields given by
Eqs.~(\ref{Brad2AntiFinal}) and (\ref{Erad2AntiFinal}), are the
first corrections to the leading order term given by
Eqs.~(\ref{BdipoloEletrico}) and (\ref{EdipoloEletrico}). In this
sense, we can write the first multipole contributions to the
multipole expansion for the radiation fields of a completely
arbitrary, but localized, time-varying source at rest in vacuum as
\begin{align}\label{Brad1final}
\textbf{B}_{\rm rad}({\bf x},t) &= \frac{1}{c^2 r}\left\{
\ddot{\textbf{p}}(t_{0}) \times\hat{\textbf{x}} + [\ddot{\bf
m}(t_0)\times\hat{\bf x}] \times\hat{\bf x} + \frac{1}{3c}
\stackrel{...}{\bf Q}(\hat{\bf x},t_0)
\times\hat{\textbf{x}}\; + \ldots\right\} \\
\textbf{E}_{\rm rad}({\bf x},t) &= \textbf{B}_{\rm rad}({\bf x},t)
\times \hat{\bf x}.
\end{align}
With these radiation fields, we can calculate the corresponding
Poynting vector and the power radiated by the source. In this paper
we have presented an alternative approach of obtaining  the first
multipole contributions to the radiation fields of general radiating
systems directly from Maxwell's equations. We hope that it  may be
useful for those who enjoy studying electromagnetism.

\begin{acknowledgments}
We are indebted to the referees for their valuable comments and
suggestions. R.S.M. and C.F. would like to thank CNPq for partial
financial support and M.M. would like to thank FAPERJ for financial
support.
\end{acknowledgments}

\end{document}